\documentclass[11pt]{scrartcl}

\usepackage{lmodern} 

\usepackage{graphicx} 
\usepackage{amsmath,amsfonts,amssymb} 
\usepackage{bm} 
\usepackage{cite}

\title{Systematic Error Estimation for Chemical Reaction Energies}
\author{Gregor N.\ Simm and Markus Reiher\thanks{corresponding author: markus.reiher@phys.chem.ethz.ch; Phone: +41446334308; Fax: +41446331594}
\vspace{10 mm}\\
ETH Z\"urich, Laboratory of Physical Chemistry, \\ Vladimir-Prelog-Weg 2, 8093 Z\"urich, Switzerland.
}

\begin{document}

\maketitle

\begin{abstract}
For the theoretical understanding of the reactivity of complex chemical systems accurate relative energies between intermediates and transition states are required.
Despite its popularity, density functional theory (DFT) often fails to provide sufficiently accurate data, especially for molecules containing transition metals.
Due to the huge number of intermediates that need to be studied for all but the simplest chemical processes, DFT is to date the only method that is computationally feasible.
Here, we present a Bayesian framework for DFT that allows for error estimation of calculated properties.
Since the optimal choice of parameters in present-day density functionals is strongly system dependent, we advocate for a system-focused re-parameterization.
While, at first sight, this approach conflicts with the first-principles character of DFT that should make it in principle system independent, we deliberately introduce system dependence because we can then assign a stochastically meaningful error to the system-dependent parametrization that makes it non-arbitrary.
By re-parameterizing a functional that was derived on a sound physical basis to a chemical system of interest we obtain a functional that yields reliable confidence intervals for reaction energies.
We demonstrate our approach at the example of catalytic nitrogen fixation.
\end{abstract}

\section{Introduction}

To understand the reactivity of a chemical system, the potential energy surface (PES) needs to explored to high accuracy.
The electronic contribution to accurate relative energies between intermediates and transition states is particularly difficult to calculate (next to the entropic contribution).
While state-of-the-art quantum chemical calculations can yield highly accurate results even for large systems \cite{Claeyssens2006_eng}, they are computationally expensive and therefore restricted to a comparatively small number of structures that can be investigated.
As a consequence, density functional theory (DFT) is currently the method of choice --- despite its shortcomings with respect to accuracy and lack of systematic improvability.
If, however, the error of each result was known, the value of approximate DFT approaches would be dramatically increased as it would flag those results to be considered with caution. An assigned error would allow one to judge whether this error compromises conclusions drawn from the data.

Most approximate exchange--correlation (XC) density functionals are constructed by fitting their parameters to benchmark data sets.
While many extensive data sets exist, such as the ones proposed by Pople \cite{Pople1989,Curtiss1991,Curtiss1997,Curtiss2000}, Truhlar \cite{Lynch2003a,Lynch2003b,Lynch2003,Zhao2004,Schultz2005a,Schultz2005,Zhao2005,Zhao2005a,Zhao2007,Zhao2008a,Zhao2009}, and Grimme \cite{Korth2009,Goerigk2010,Goerigk2011}, studies have shown that the accuracy of XC functionals can be strongly system dependent \cite{Curtiss1997,Curtiss2000,Salomon2002,Zhao2004,Curtiss2005,Riley2007,Weymuth2014}, which, naturally, will become more severe for short-lived reactive intermediates.

Moreover, it is not certain that the accuracy reported in benchmark studies is transferable to a specific system under consideration.
For instance, many benchmark data sets contain transition metals \cite{Niu2000,Schultz2005a,Schultz2005,Furche2006,Jiang2012,Weymuth2014}, however, most of them include only small (unsaturated and thus atypical) compounds (e.g., transition metal dimers).
Electronic structures exhibited by transition-metal complexes are so diverse that it is very difficult to represent them in an unbiased benchmark set.
If accurate reference data for the chemical system of interest were available, one could not even assume the error of a DFT result to be constant among homologous molecules \cite{Weymuth2014,HallHyd}.
In a recent study \cite{Weymuth2014}, we showed that popular density functionals struggle to reproduce experimental ligand dissociation energies of large organometallic transition-metal complexes in our WCCR10 reference set.
Moreover, the results obtained with different popular density functionals deviate significantly from one another in an irregular manner \cite{Weymuth2014}.
However, we also showed that they can be re-parametrized to yield exactly the reference energies \cite{Weymuth2015}.
This indicates that the parameters of the standard functional investigated in Ref.~\cite{Weymuth2015} are flexible enough 
to be chosen to exactly reproduce all coordination energies of the WCCR10 set. 
There exists, however, no unique parameter set that is equally accurate for all WCCR10 coordination energies at the same time.

It is, therefore, difficult to predict the accuracy of density functional calculations in general.
It is common practice \cite{cramerbook} (see also benchmark studies such as the one in Ref.~\cite{Goerigk2011a}) to investigate the spread of results from a selection of present-day density functionals to estimate the sensitivity of the investigated property with respect to functional form and choice of parameters.
But as the selection of functionals is in parts arbitrary, this approach is highly unsystematic and the spread has no statistical significance.
Therefore, a systematic framework for the assessment of accuracy of density functionals is required.

In 2005, N{\o}rskov, Sethna, Jacobsen, and co-workers presented a scheme for systematic error estimation of DFT results \cite{Mortensen2005} based on Bayesian statistics \cite{Jaynes2003,Gelman2013} (see also Refs.\ \cite{Brown2003,Frederiksen2004,Petzold2012}).
In their approach, an ensemble of XC functionals is generated by which a mean and a variance can be assigned to each computational result.
Two types of density functionals were designed within this framework: BEEF-vdW \cite{Wellendorff2012} and mBEEF \cite{Wellendorff2014,Pandey2015}.
While both functionals were parameterized employing a wide range of data sets \cite{Wellendorff2012,Wellendorff2014}, transition metal complexes were not 
included and also transferability issues remain (especially for such complexes). 
In addition, BEEF-vdW and mBEEF are both pure functionals, whereas, it is well known that hybrid functionals tend to be more accurate than pure functionals (see, e.g., Refs.~\cite{Riley2007,Weymuth2014}).
Along these lines, Zabaras and coworkers \cite{Aldegunde} developed a new exchange-correlation functional employing a Bayesian approach combined with machine learning to predict bulk properties of transition metals and monovalent semiconductors.
Very recently, Vlachos and coworkers successfully applied Bayesian statistics to DFT reaction rates on surfaces \cite{Sutton2016}.
However, so far the application of Bayesian statistics in DFT has been limited to solid-state and surface chemistry \cite{Gautier2015}.

Here, we develop Bayesian error estimation for molecules.
It is one goal of this study to obtain a class of hybrid functionals that accurately describes the reaction energies of a specific chemical system.
We advocate for a system-focused re-parametrization of our ensemble of density functionals to overcome the issue of transferability, while preserving standard design principles of density functionals.
Through Bayesian statistics, our class of functionals reports uncertainties for each calculated result which eliminates the arbitrariness of a system-specific parametrization.

We demonstrate our approach at the first example of synthetic catalytic nitrogen fixation under ambient conditions: the Chatt-Schrock cycle \cite{Chatt1978,Yandulov2002,Yandulov2003}.
Recently, we presented alternative pathways of this catalytic cycle \cite{Bergeler2015}.
To reliably assess the relevance of such alternative catalytic pathways, confidence intervals for reaction energies and barriers are a mandatory prerequisite and can be obtained from Bayesian error estimation.

\section{Theory}

\subsection{Error Estimation in DFT}\label{subsec:bee}

The parameters $\bm{a}$ of a density functional are usually determined by parametrization to some data set 
$\mathcal{D}=\{(i,\mathcal{R}(i))\}$ 
containing molecular structures $i$ and an observerable which is determined with a (experimental or computational) reference method $\mathcal{R}$
(with the exception of those fixed by exact DFT conditions)
This is accomplished by minimizing a cost function $C(\bm{a})$ to obtain a best fit $\bm{a}_0$, which is then reported.
However, information on the neighborhood of $C(\bm{a}_0)$ is thereby lost.
For instance, it cannot be determined if the reported minimum is shallow or steep (see Ref.~\cite{Weymuth2015}) or how perturbations in the parameter space (e.g., due to a new item in the data set) translate into variations of some observable $\mathcal{O}$.

Instead of considering only the best-fit parameters, one can assign a conditional probability distribution to the continuous set of parameters 
\begin{equation}
  p_a = p(a | \mathcal{O}, D) \propto \exp\left(- \frac{C(a)}{T} \right),
\end{equation}
where the observable $\mathcal{O}$ is obtained from a single linear parameter $a$, and $C$ denotes 
a cost function quadratic in $a$ \cite{Gelman2013,Jaynes2003}.
It can be shown \cite{Mortensen2005} that the spread of this distribution is determined by the ensemble temperature $T = 2C(a_0)$ (see Eq.~(\ref{eq:temp}) below).
A standard parametrization of density functionals can be considered a special case of this distribution where $T = 0$, so that $p(a | \mathcal{O}, D) = \delta (a - a_0)$ \cite{Frederiksen2004,Mortensen2005,Petzold2012}.

In practice, this distribution needs to be sampled for which a set of parameters $\{a_1, a_2, ..., a_N\}$ is generated.
It can be shown \cite{Petzold2012} that, with a cost function quadratic in $a$, a Gaussian distribution $\mathcal{N}$,
\begin{equation}\label{eq:normal_distribution}
  p_a = \mathcal{N}(a_0, \sigma^2),
\end{equation}
with mean $a_0$ and variance $\sigma^2 = T / (\partial^2 C(a) / \partial a^2 |_{a_0})$ must be sampled.
From the ensemble of parameters, a confidence interval can be calculated for any observable $\mathcal{O}$ \cite{Mortensen2005}.

\subsection{Brief Derivation of Error Estimation for DFT}\label{subsec:deriv}
Consider some observable $\mathcal{O}^{\bm{a}}$ with parameters $\bm{a}$ to be calculated for some molecular system $i$.
In this work, the observable will be the energy difference between a pair of structural isomers.
We now approximate a reference result $\mathcal{R}(i)$ for system $i$ by $\mathcal{O}^{\bm{a}}$ and therefore define
\begin{equation}
  \Delta^{\bm{a}}(i) = \mathcal{O}^{\bm{a}}(i) - \mathcal{R}(i).
\end{equation}
We aim to find a probability distribution $p_{\bm{a}}$ so that, across the data set $\mathcal{D}$, the deviation of $\mathcal{O}^{\bm{a}}$ 
from $\mathcal{O}^{\bm{a}_0}$, 
\begin{equation}\label{deltadef}
\delta^{\bm{a}}(i) = \mathcal{O}^{\bm{a}}(i) - \mathcal{O}^{\bm{a}_0}(i),
\end{equation}
is, on average, equal to the deviation of $\mathcal{O}^{\bm{a}}$ from $\mathcal{R}$, i.e.:
\begin{equation}\label{eq:condition0}
   \sum_{i \in \mathcal{D}} \left\langle [\delta^{\bm{a}}(i)]^2 \right\rangle_{\bm{a}} = \sum_{i \in \mathcal{D}} [\Delta^{\bm{a}_0}(i)]^2,
\end{equation}
where $\bm{a}_0$ is the parameter set that minimizes the cost function $C(\bm{a})$,
\begin{equation}\label{Cdef}
C(\bm{a}) = \sum_{i \in \mathcal{D}} [\Delta^{\bm{a}}(i)]^2.
\end{equation}
Defining the quadratic deviation of a parameter set $\bm{a}$ from the optimal set $\bm{a}_0$ as $F(\bm{a})$,
\begin{equation}\label{Fdef}
F(\bm{a}) = \sum_{i \in \mathcal{D}} [\delta^{\bm{a}}(i)]^2,
\end{equation}
we can write Eq.\ (\ref{eq:condition0}) in more compact form as
\begin{equation}\label{eq:condition}
\left\langle F(\bm{a}) \right\rangle_{\bm{a}} = C(\bm{a}_0)
\end{equation}

To obtain the probability distribution with the highest information entropy, we maximize the Shannon entropy of the distribution under the condition in Eq.~(\ref{eq:condition}).
Introducing a fixed number $N$ of parameter sets $\{\bm{a}_k \}$ and obeying that the sum over all probabilities equals one as an additional constraint, 
we have for the variation of the resulting Lagrangian function with respect to the probability $p_{\bm{a}_j}$ of one of these parameter sets $\bm{a}_j$
\begin{equation}\label{eq:lagrange}
  \frac{\partial}{\partial p_{\bm{a}_j}} \left( - \sum^N_{k=1} p_{\bm{a}_k} \ln(p_{\bm{a}_k}) - \lambda \left( C(\bm{a}_0) - \sum_{k=1}^N p_{\bm{a}_k} F({\bm{a}_k)} \right) - \mu \left(1 - \sum_{k=1}^N p_{\bm{a}_k} \right) \right) \stackrel{!}{=} 0,
\end{equation}
where $\lambda$ and $\mu$ are Lagrange multipliers.
Solving Eq.~(\ref{eq:lagrange}) yields the well-known relation
\begin{equation}\label{eq:pdf}
  p_{\bm{a}_j} = \frac{\exp( - \lambda F(\bm{a}_j))}{\sum_{k=1}^N \exp (- \lambda F(\bm{a}_k))}.
\end{equation}

To determine the Lagrange multiplier $\lambda$, we consider an observable $\mathcal{O}^a$ with a single linear parameter $a$,
\begin{equation}\label{Olineara}
\mathcal{O}^a(i) = a x_i + c.
\end{equation}
Then $F(\bm{a})$ simplifies to
\begin{equation}\label{eq:F}
  F(a) = \sum_{i \in \mathcal{D}} ((a - a_0) \cdot x_i)^2.
\end{equation}
The expectation value of $F(a)$ for the $N$ parameters $\{a_k \}$ can be written as
\begin{equation}\label{eq:harmonic}
  \langle F(a) \rangle_{a_k} = \frac{\sum_{k=1}^N F(a_k) \exp(-\lambda F(a_k))}{\sum_{k=1}^N \exp(-\lambda F(a_k))}.
\end{equation}
According to the equipartition theorem, each harmonic degree of freedom contributes $T/2$ to the cost (with the Boltzmann constant taken to be one), 
which implies for Eq.~(\ref{eq:condition}) in our single-parameter model that
\begin{equation}\label{eq:temp}
  \langle F(a) \rangle_{a_k} = C(a_0) = \frac{1}{2} T ,
\end{equation}
so that an expression for $\lambda$ which corresponds to the inverse ensemble temperature $T$, can be derived \cite{Frederiksen2004,Mortensen2005,Petzold2012}.

Finally, the probability distribution $p_a$ needs to be sampled.
From the definition of $C(a)$ we have for a single linear parameter
\begin{equation}\label{eq:C_simple}
C(a) = \sum_{i \in \mathcal{D}} \left[ (ax_i+c_0) - (a_0x_i+c_0) \right]^2
\end{equation}
and may expand $C(a)$ around $C(a_0)$
\begin{equation}\label{eq:C_expand}
C(a)=C(a_0)+\frac{1}{2} \left.\frac{\partial^2 C(a)}{\partial a^2}\right|_{a_0} (a - a_0)^2 + \cdots~.
\end{equation}
The second derivative of $C(a)$ at the position $a=a_0$ is easy to evaluate
\begin{equation}\label{eq:F_alt}
\left.\frac{\partial^2 C(a)}{\partial a^2}\right|_{a_0} = \sum_{i \in \mathcal{D}} 2x_i^2
\end{equation}
so that with Eq.~(\ref{eq:F}) and Eq.~(\ref{eq:C_expand}) we find
\begin{equation}\label{eq:F_alt}
  F(a) = \frac{1}{2} \left.\frac{\partial^2 C(a)}{\partial a^2}\right|_{a_0} (a - a_0)^2 .
\end{equation}
From Eqs.~(\ref{eq:pdf}) and (\ref{eq:F_alt}), it can be seen that the probability distribution of $a$ is a normal distribution: 
\begin{equation}
\label{normaldis}
  p_a  = \mathcal{N} \left(a_0, T \left/ \left.\frac{\partial^2 C(a)}{\partial a^2}\right|_{a_0} \right. \right).
\end{equation}
This distribution is then sampled by choosing the parameters $\{a_k \}$ of the $N$ models (the samples) so that a standard deviation $\sigma$ for 
the observable $\mathcal{O}$ of system $i$ can be calculated
\begin{equation}\label{eq:standard_deviation}
  \sigma(\mathcal{O}(i)) = \sqrt{ \frac{1}{N} \sum_{k=1}^N \Big(\mathcal{O}^{a_k}(i) - \mathcal{O}^{a_0}(i) \Big)^2 }.
\end{equation}
$N$ must be chosen such that $\sigma(\mathcal{O}(i))$ is converged. The sets of linear parameters $\bm{a}_k$ (or $a_k$ in the case of a single 
linear parameter) are obtained from computer generated random numbers with the normal distribution in Eq.\ (\ref{normaldis}).

\subsection{Range Separation}\label{subsec:range}

In this study, the parameters of the range-separated hybrid (RSH) version of the popular density functional PBE0 \cite{Perdew1996a,Adamo1999,Perdew1996} are considered for Bayesian error estimation for the following reasons:
Firstly, exact exchange plays an important role in the description of transition metals \cite{Reiher2001,Salomon2002,Reiher2002,Riley2007,Weymuth2014}.
Secondly, many issues of present-day density functionals, such as the underestimation of barriers of chemical reactions, can be attributed to the delocalization error \cite{Cohen2008}.
Baer et al.\ showed that long-range corrected (LC) functionals appear to have resolved this issue \cite{Baer2010}.
Finally, it was observed \cite{Peach2006,Vydrov2006,Rohrdanz2008,Stein2009,Srebro2012,Srebro2012a,Autschbach2014} that the parameters in the RSH scheme are in fact system dependent and that their adjustment can improve the functional's accuracy.

In RSH functionals \cite{Leininger1997,Iikura2001,Heyd2003,Tawada2004,Yanai2004,Arbuznikov2012}, the exchange functional is divided into short-range DFT exchange and long-range Hartree-Fock (HF) exchange by splitting the electron-electron interaction operator $1/r_{12}$:
\begin{equation}\label{eq:lc}
  \frac{1}{r_{12}} = \underbrace{\frac{1 - [\alpha + \beta \cdot \text{erf}(\gamma r_{12})]}{r_{12}}}_\text{short-range} + 
             \underbrace{\frac{\alpha + \beta \cdot \text{erf}(\gamma r_{12})}{r_{12}}}_\text{long-range}
\end{equation}
This ansatz introduces three adjustable parameters: $\alpha$, $\beta$, and the range-separation parameter $\gamma$.
In the long-range corrected scheme, only two are independent since $\alpha + \beta = 1$ if the two operators on the right-hand side
of Eq.\ (\ref{eq:lc}) are evaluated by different energy expressions.
LC-PBE0 is such a functional, where $\alpha = 0.25$, $\beta = 0.75$, and $\gamma = 0.3$ (if $\alpha = 0.25$, $\beta = 0.75$, and $\gamma = 0$, PBE0 \cite{Adamo1999} is recovered).
By contrast, in the Coulomb-attenuating method by Yanai et al.\ \cite{Yanai2004}, $\alpha = 0.19$, $\beta = 0.46$, and $\gamma = 0.33$, so that $\alpha + \beta = 0.65$.
However, only for $\alpha + \beta = 1$ the potential shows the correct asymptotic behavior of $1/r_{12}$ \cite{Srebro2012}.

\subsection{Parameters in PBE}

In addition to the parameters in the LC scheme, we optimize parameters of the original PBE functional \cite{Perdew1996a} to increase model flexibility.
In Hartree atomic units, the correlation part of the PBE functional can be written as
\begin{equation}
  E_c^\text{PBE}[\rho_\uparrow, \rho_\downarrow] = \int \rho \left[ \epsilon_c^\text{unif} (r_s, \zeta) + H(r_s, \zeta, t) \right] \; d^3 r,
\end{equation}
with
\begin{equation}
  H(r_s, \zeta, t) = \gamma_c \phi^3 \ln \left( 1 + \frac{\beta_c}{\gamma_c} \frac{t^2 + A t^4}{1 + A t^2 + A^2 t^4} \right),
\end{equation}
where $\rho = \rho_\uparrow + \rho_\downarrow$ is the electron density (obtained as a sum of spin-up and spin-down densities), 
$\epsilon_c^\text{unif}(r_s, \zeta)$ the correlation energy per particle of the uniform electron gas,
$r_s = \left[ (4 \pi/ 3) \rho \right]^{1/3}$ the local Wigner-Seitz radius,
$t = |\nabla \rho| / (2 \phi k_s \rho)$ the correlation density gradient,
$\zeta = (\rho_\uparrow - \rho_\downarrow) / \rho$ the relative spin polarization,
and $\phi = ((1 + \zeta)^{2/3} + (1 - \zeta)^{2/3}) / 2$ a spin scaling factor.
The factor $A$ is a function of $\phi$ and $\epsilon_c^\text{unif}$ \cite{Perdew1996a}.
The parameter $\beta_c = 0.066725$ is the second-order gradient expansion coefficient of the correlation energy in the high-density limit and the parameter $\gamma_c = (1 - \ln 2) / \pi^2$ is given by the uniform scaling to the high-density limit of the spin-unpolarized correlation energy.

The exchange part of the PBE functional is given by
\begin{equation}
  E_x^\text{PBE}[\rho] = \int \rho \, \epsilon_x^\text{unif}(\rho) F_x^\text{PBE}(s) \; d^3 r,
\end{equation}
where $F_x^\text{PBE}(s) = 1 + \kappa - \kappa / (1 + \frac{\mu}{\kappa} s^2)$, $\kappa = 0.804$, and the reduced gradient $s = |\nabla \rho| / (2 k_F \rho)$.
The parameter $\kappa$ is determined by the Lieb--Oxford bound \cite{Lieb1981} for the exchange energy, and the parameter $\mu$ is determined to satisfy the correct linear response of the spin-unpolarized uniform electron gas ($\mu = \beta_c \pi^2/3$) such that $\mu = 0.21951$.

Since its introduction, many variations of the original PBE functional were presented, such as revPBE \cite{Zhang1998}, PBEsol \cite{Perdew2008,Perdew2009}, and APBE \cite{Constantin2011}.
In these functionals, the functional form of PBE is kept, however, the parameters $\mu$, $\beta_c$, and $\kappa$ are varied.
A study by Della Sala and coworkers \cite{Fabiano2011} showed that a property-specific optimization of these parameters can lead to an increase in accuracy.

\subsection{Model Definition}\label{subsec:modeldef}
In this study, we adjust the parameters $\alpha$, $\gamma$, $\mu$, and $\kappa$ to obtain a class of functionals LC$^\star$-PBE0($\mathcal{D}$)
that allows us to describe a particular system of interest represented by reference data $\mathcal{D}$;
for this optimization we choose the L-BFGS-B scheme \cite{Byrd1995}.
Although this system-specific parametrization is generally viewed as an illicit
departure from the first-principles character of DFT toward a semi-emipirical approach \cite{Burke2012}, it is key to accurate error estimation in this work.
A small number of parameters comes with the advantage that a small data set suffices for the parametrization.
Being the only parameter that contributes linearly to the total electronic energy, $\alpha$ is then considered in the error estimation
protocol, keeping the other parameters constant at their re-optimized value.
We wish to emphasize that the linearity of the energy with respect to $\alpha$ will only be guaranteed 
if the energies are calculated non-selfconsistently, i.e., employing the same electron density.
In this work, we calculate the electronic energy of the ensemble non-self-consistently employing the electron density obtained from a 
self-consistent calculation with the best-fit parameters $a_0$ \cite{Wellendorff2012,Wellendorff2014}.
Therefore, the error estimation scheme does not result in a significant computational overhead.
In the Supporting Information, we show that this approximation can be well justified.

\subsection{Reference Data}\label{subsec:dataset}

For an accurate re-parametrization, the reference data set needs to be representative for the system of interest.
Specifically, the data set should contain structures that are intermediates and transition states of the chemical process under consideration.
Of course, one cannot expect to include every relevant structure, but the stochastic nature of our approach takes this limitation into account.
Moreover, knowledge-based Bayesian statistics may even be considered in a rolling re-parametrization scheme, in which more accurate reference data are constantly added when they become available.

\begin{figure}
\centering
\includegraphics[width=0.7\textwidth]{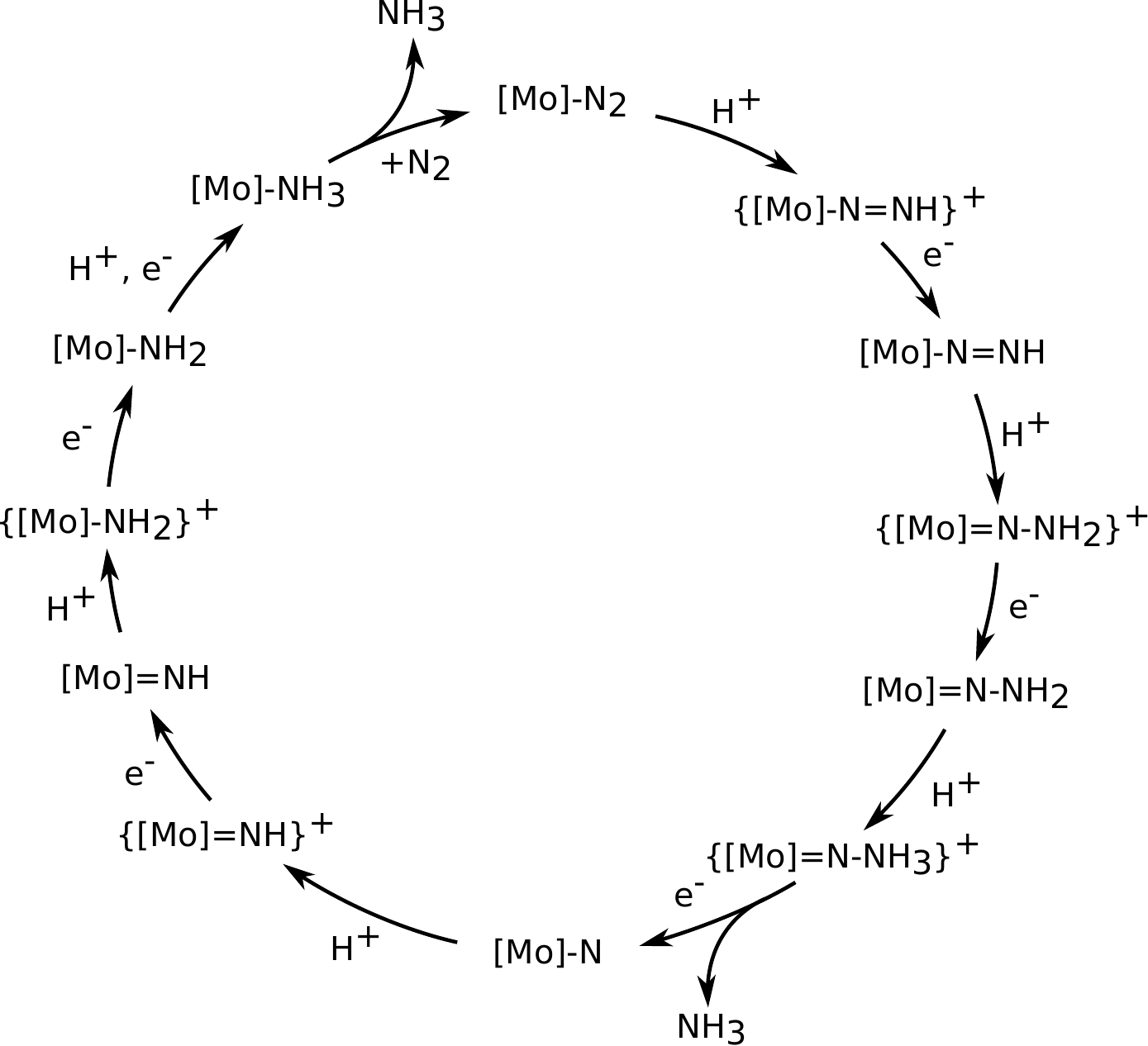}
\caption{Chatt--Schrock nitrogen-fixation cycle.}
\label{fig:cycle}
\end{figure}

In this study, the chemical reactivity of the catalyst synthesized by Yandulov and Schrock \cite{Yandulov2002,Yandulov2003} is investigated.
A proposed catalytic cycle for this catalyst is the Chatt--Schrock cycle \cite{Chatt1978,Yandulov2002,Yandulov2003}, in which intermediates are formed by a 
sequence of protonation and reduction steps (see Fig.~\ref{fig:cycle}).
The acid 2,6-lutidinium (LutH) and reducing agent decamethylchromocene (CrCp$^*_2$) are the sources of protons and electrons, respectively.
We have investigated this system in great detail in the past decade \cite{LeGuennic2005,Reiher2005,Schenk2008,Schenk2009,Schenk2009a,Bergeler2015b,Bergeler2015}.
In a recent study \cite{Bergeler2015}, we showed that numerous relevant isomers of Schrock intermediates are likely to be formed by protonation and reduction alone.

If only little experimental reference data exists for a chosen system, highly accurate post-HF methods, such as coupled-cluster theory, can be employed.
Usually, their steep scaling of computing time with system size require the restriction to rather small model systems.

\begin{figure}
\centering
\includegraphics[width=0.8\textwidth]{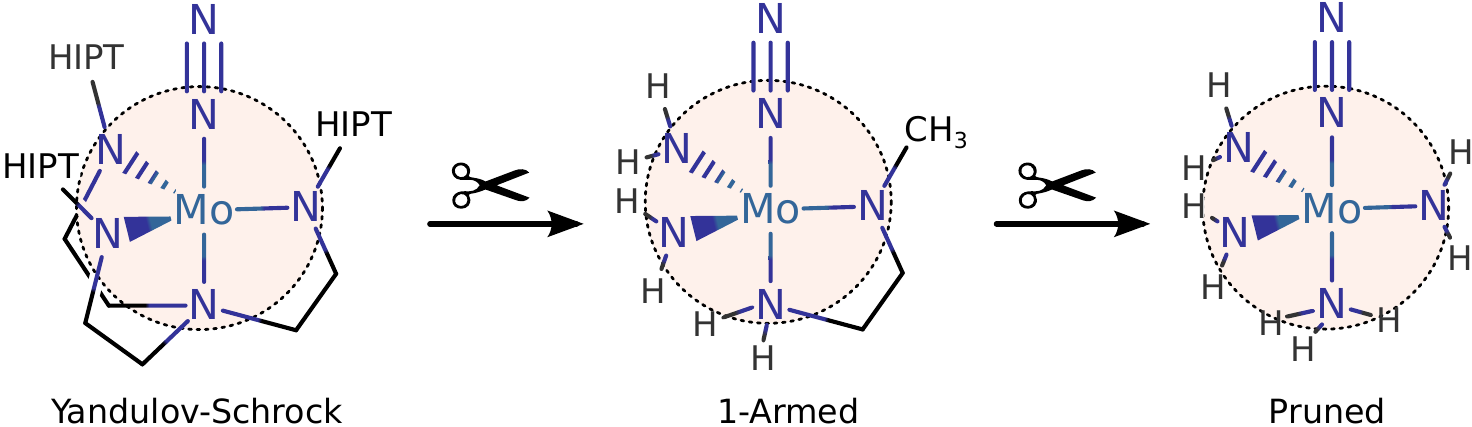}
\caption{Model systems for the Yandulov--Schrock catalyst.
While keeping the first coordination sphere (dashed circle) intact, carbon and hydrogen atoms are removed to reduce computational effort.
}
\label{fig:pruning}
\end{figure}

\begin{figure}
\centering
\includegraphics[width=0.8\textwidth]{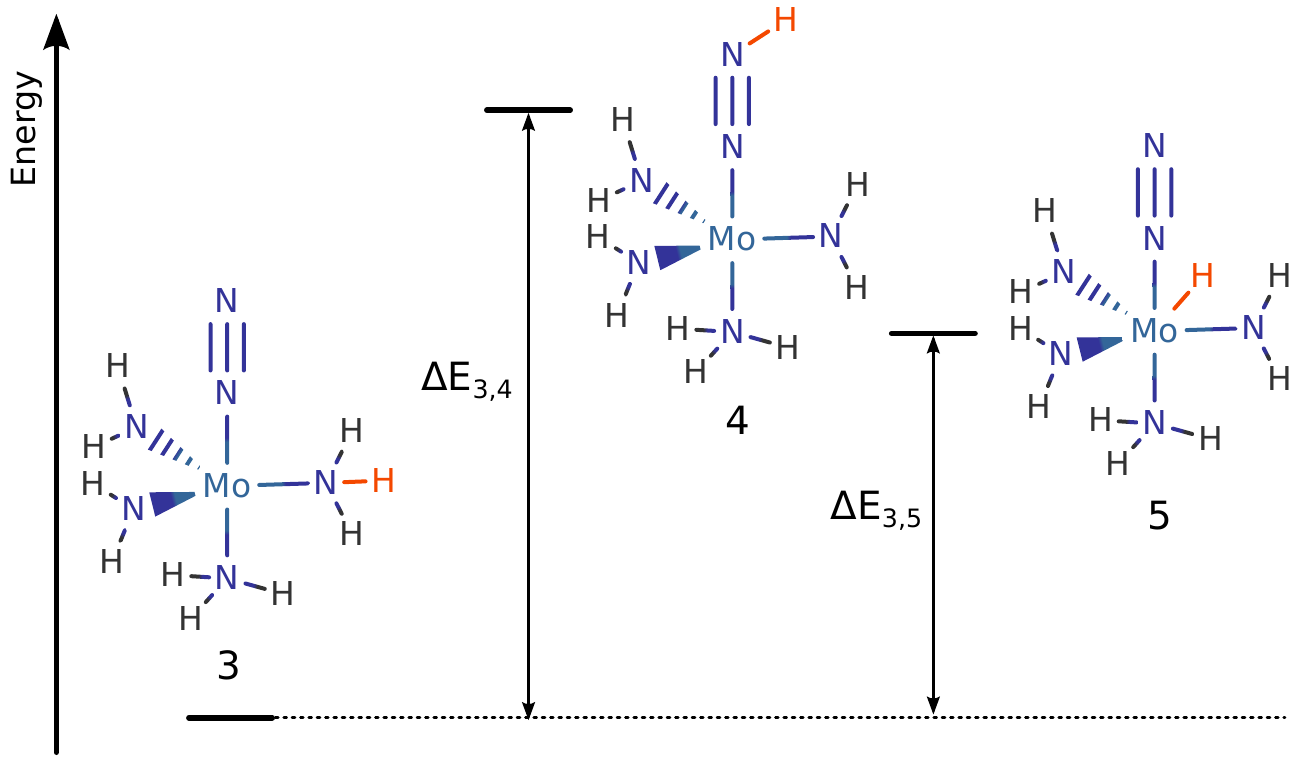}
\caption{Example for relative energies $\Delta E_{3,4}$ and $\Delta E_{3,5}$ between three isomers (structures 3, 4, and 5 in $D_\text{P}$) of the pruned Yandulov--Schrock complex.}
\label{fig:relative_energies}
\end{figure}

For the construction of the reference data set, we chose the CCSD(T) method; i.e., $\mathcal{R}$ is CCSD(T).
Moreover, a model is constructed in which the hexa-\textit{iso}-propyl terphenyl (HIPT) substituents are replaced by methyl groups or hydrogen atoms; in this way the computational effort is reduced, while the first coordination sphere remains intact (see Fig.~\ref{fig:pruning}).
To probe the transferability of our functional optimized on data for the (\textit{pruned}) model system to the original complex, 
an intermediate (\textit{1-armed}) model is also investigated.
The resulting reference data sets, referred to as $D_\text{P}$ and $D_\text{A}$, accordingly, contain energy differences between structures on the same PES, i.e., structures with the same number and type of atomic nuclei, the same number of electrons, and the same electronic spin state (see Fig.~\ref{fig:relative_energies} for an example of two reference values).
The structure coordinates and reference electronic energies of all structures considered in this study are given in the Supporting Information.

The observable $\mathcal{O}$ is the energy difference $\Delta E_{i,j}$ between the structural isomers $i$ and $j$.
Then, the cost function $C$ employed in the parametrization reads
\begin{equation}
  C(\alpha, \gamma, \kappa, \mu) = \sum_{i,j \in D} \left( \Delta E_{i,j} (\alpha, \gamma, \kappa, \mu) - \Delta E_{i,j}^\text{ref} \right)^2 = \sum_{i,j \in D} C_{i,j}(\alpha, \gamma, \kappa, \mu),
\end{equation}
where $\Delta E_{i,j} (\alpha, \gamma, \kappa, \mu)$ and $\Delta E_{i,j}^\text{ref}$ are the relative energies obtained with the LC-PBE0 functional with parameters ($\alpha$, $\gamma$, $\kappa$, $\mu$) and the reference value, respectively, and $i$ and $j$ are structures on the same PES.

\subsection{Computational Methodology}\label{subsec:comp}

All BP86/RI/def2-TZVP \cite{Perdew1986,Becke1988,Weigend2005} model-catalyst structures in $D_\text{P}$ and $D_\text{A}$ were optimized with the program package \textsc{Turbomole} \cite{turbomole2013}.
BP86/RI/TZVP+SV(P) optimized structures of the full Yandulov--Schrock catalyst were taken from Ref.~\cite{Schenk2008}.

All CCSD(T) single-point calculations were carried out with the \textsc{Molpro} 2010.1 \cite{Werner2012,molpro2010_1} program package.
For the elements hydrogen, carbon, and nitrogen the aug-cc-pVDZ basis set \cite{Woon1994} was chosen.
For molybdenum a double-$\zeta$ basis set together with an effective core potential (aug-cc-pVDZ-PP) was employed \cite{Peterson2007}.
Clearly, for truly accurate reference data much larger one-electron basis sets or F12 basis sets are required.
However, we already stress at this point that all conclusions drawn in this work will remain unchanged if the reference energies are corrected by a constant energy shift that may be different for different pairs of structures.

All subsequent DFT single-point calculations were carried out with the NWChem program package \cite{Valiev2010}.
The following density functionals were employed: BP86 \cite{Perdew1986,Becke1988}, B3LYP \cite{Lee1988,Becke1993,Becke1988}, PBE \cite{Perdew1996a}, PBE0 \cite{Adamo1999}, LC-PBE0, M06-2X \cite{Zhao2007}, M06-L \cite{Zhao2006}, TPSS \cite{Tao2003}, and TPSSh \cite{Staroverov2003}.
Furthermore, for BP86, B3LYP, PBE0, M06-2X, M06-L, TPSS, and TPSSh we considered Grimme's third generation dispersion correction \cite{Grimme2010,Goerigk2011a}, denoted as BP86-D3, B3LYP-D3, PBE0-D3, M06-2X-D3, M06-L-D3, TPSS-D3, and TPSSh-D3, respectively.
For all DFT calculations on structures in $D_\text{P}$ and $D_\text{A}$ a triple-$\zeta$ basis set (def2-TZVP) was chosen for all atoms \cite{Weigend2005}.
Calculations on the Yandulov--Schrock catalyst were carried out with a triple-$\zeta$ basis set (def2-TZVP) on molybdenum and nitrogen atoms, and a double-$\zeta$ basis set (def2-SV(P) \cite{Weigend2005}) on carbon and hydrogen atoms.
In all DFT calculations, scalar-relativistic effects were taken into account for the elements molybdenum and chromium by means of Stuttgart effective core potentials \cite{Andrae1990}.

Data analysis and visualization were carried out with the software packages Pandas \cite{McKinney2012}, Matplotlib \cite{Hunter2007}, and IPython \cite{Perez2007}.

\section{Results}

\subsection{Parameter Selection and Optimization}

\begin{figure}
\centering
\includegraphics[width=0.9\textwidth]{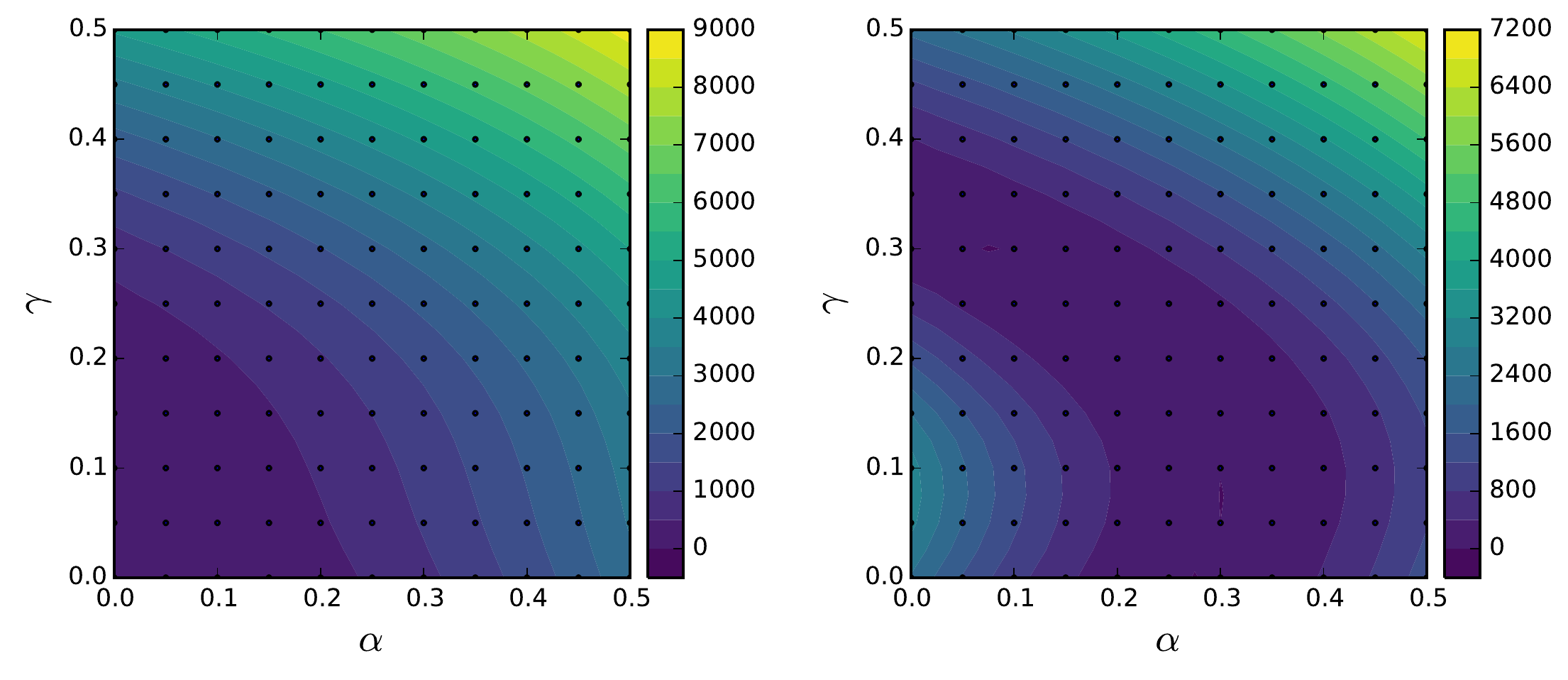}
\caption{Two cost functions, $C_{26,27}$ (left) and $C_{26,28}$ (right), depending on the parameters $\alpha$ and $\gamma$ (in (kJ/mol)$^2$).
The cost functions were calculated from the relative energies between three isomers of [Mo]-NH$_2^+$.
The parameters $\kappa=\kappa^\text{PBE}$ and $\mu=\mu^\text{PBE}$ were kept constant.
}
\label{fig:para_scan}
\end{figure}

Since the parameters $\kappa$ and $\mu$ in the PBE functional were determined by fulfilling exact boundary conditions \cite{Perdew1996a}, we first investigated whether the optimization of the parameters in the range-separation scheme, i.e., $\alpha$ and $\gamma$, suffices to obtain an accurate functional.
Accordingly, $C_{i,j}(\alpha, \gamma, \kappa=\kappa^\text{PBE}, \mu=\mu^\text{PBE})$ were calculated for structures in $D_\text{P}$ as a function of $\alpha$ ($\beta = 1 - \alpha$) and $\gamma$, whereby $\kappa$ and $\mu$ were kept constant.
As an example, the results for two relative energies between three isomers of [Mo]-NH$_2^+$ are shown in Fig.~\ref{fig:para_scan}.
Results for additional structures are given in the Supporting Information. 
Even though the three structures are similar (differing in the position of only one hydrogen atom), the optimal parameters deviate significantly (as can be seen from Fig.~\ref{fig:para_scan}).
We note that the shape of the contour plot would not change significantly for a shifted reference energy $\Delta E_i$.
A slightly different reference energy would only result in a shift of the observed pattern.
Hence, it is not decisive for this study whether or not our coupled-cluster reference data is of ultimate accuracy.

\begin{figure}
\centering
\includegraphics[width=0.9\textwidth]{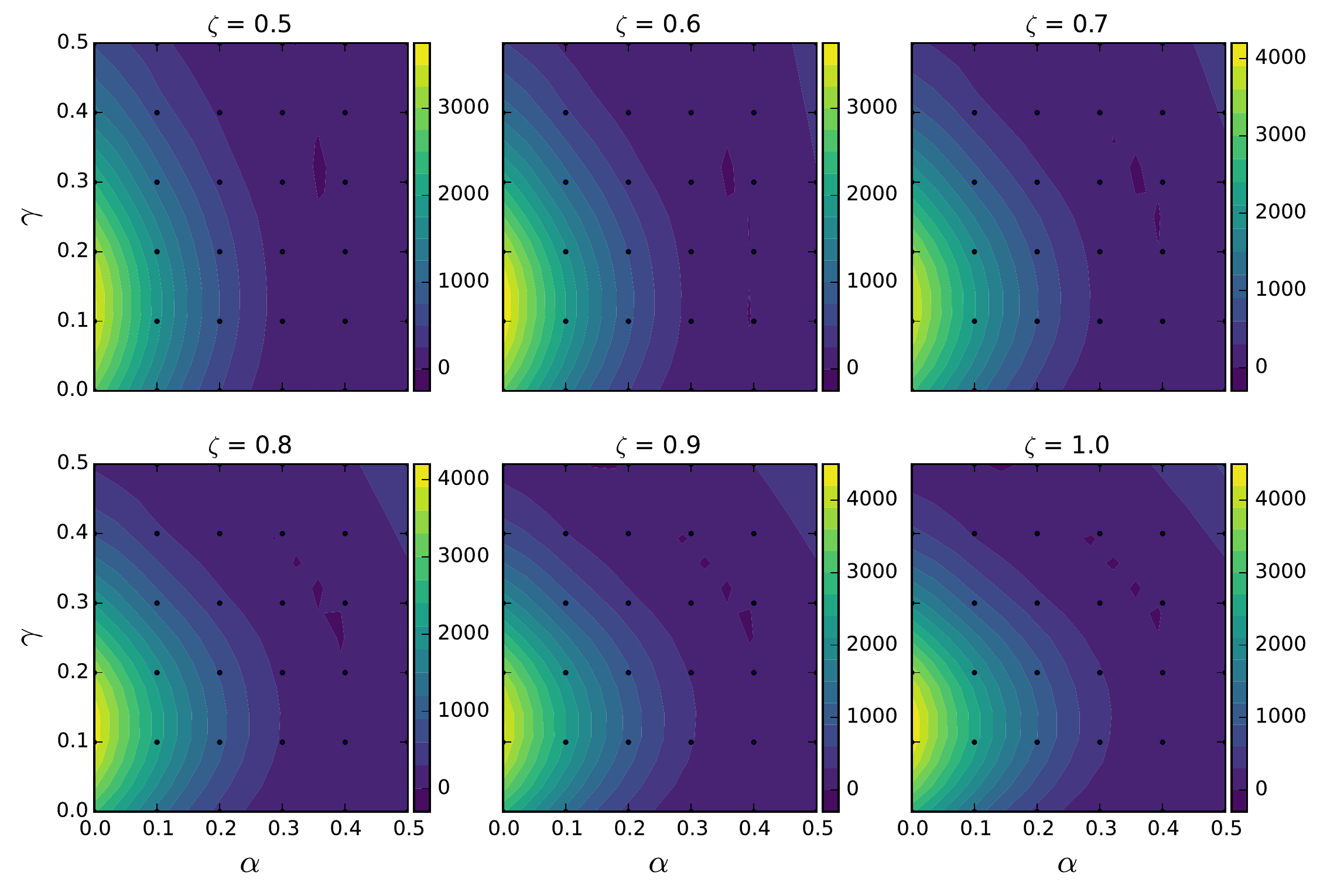}
\caption{Cost function $C_{8,11}$ depending on $\alpha$ and $\gamma$ and on the amount of long-range correction $\zeta = \alpha + \beta$ (in (kJ/mol)$^2$).
The parameters $\kappa$ 
and $\mu$   
were kept constant at their original values in the PBE functional.
}
\label{fig:cam}
\end{figure}

Furthermore, we investigated whether incomplete LC, i.e., $\alpha + \beta < 1$, can increase model flexibility.
In Fig.~\ref{fig:cam}, the amount of LC, $\zeta = \alpha + \beta$, is varied for the cost function $C_{8,11}$. 
It can be seen that the form of the contour plot is hardly affected by $\zeta$; only the curvature of the contour lines increases.
This can be understood when appreciating that the effect of $\gamma$ increases with $\zeta$ (see Eq.~(\ref{eq:lc})).
Therefore, we consider it unlikely that changing the amount of LC leads to an increase in accuracy worth compromising the correct asymptotic behavior.
For the rest of this study, we therefore preserve complete LC, i.e., $\alpha + \beta = 1$.

\begin{figure}
\centering
\includegraphics[width=0.5\textwidth]{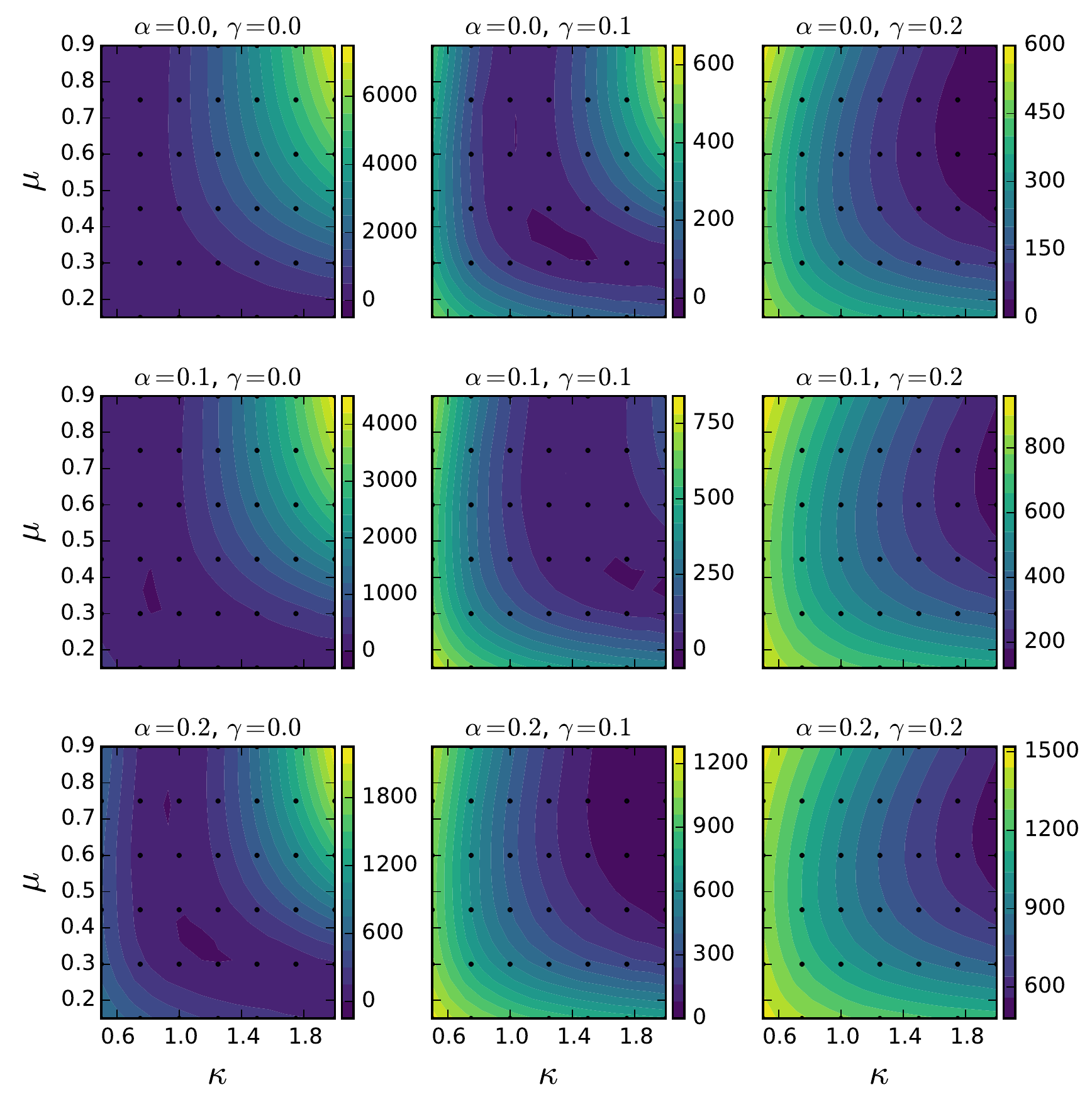}
\includegraphics[width=0.5\textwidth]{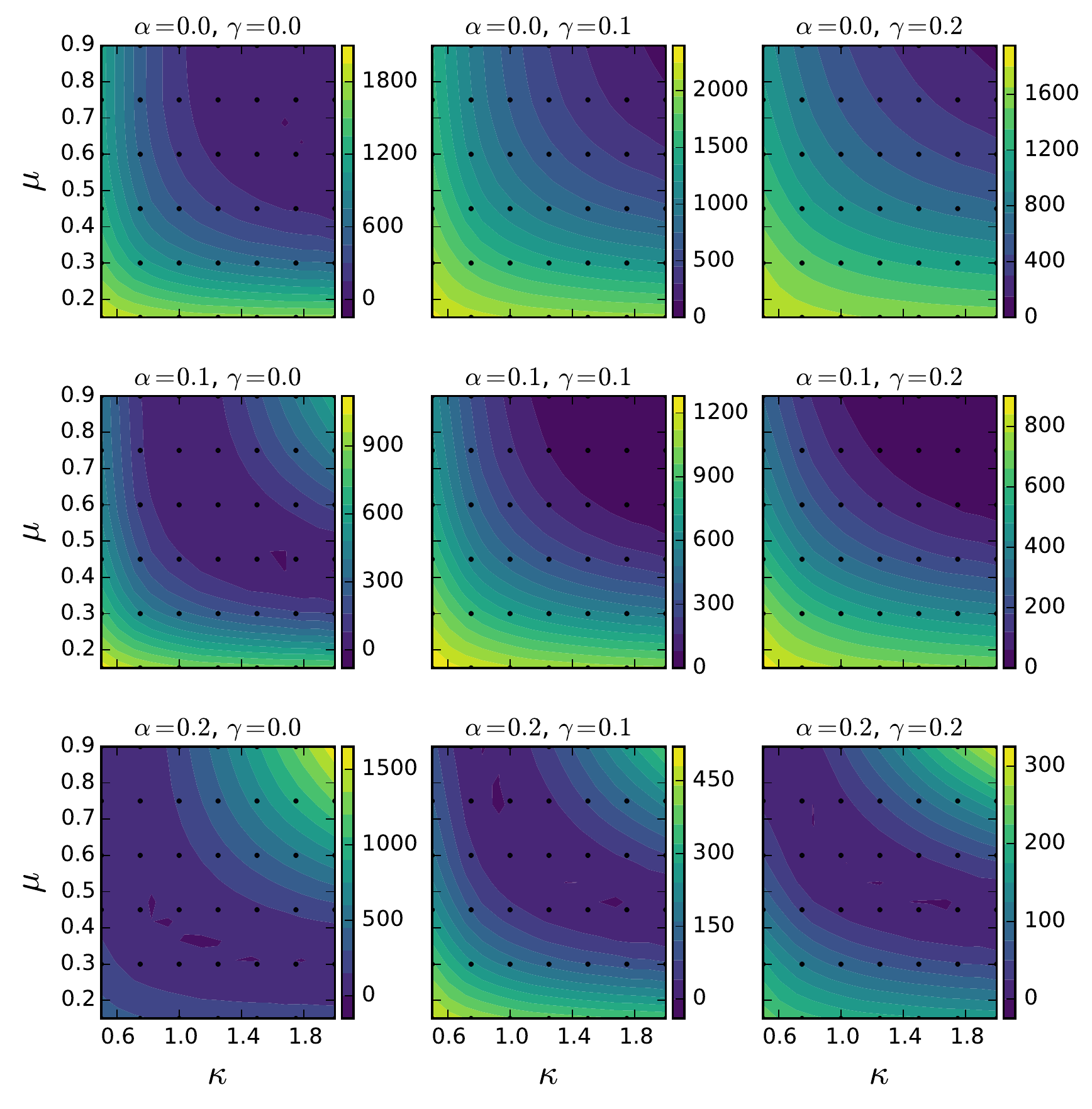}
\caption{
Cost functions $C_{23,24}$ (top 9 diagrams) and $C_{23,25}$ (bottom 9 diagrams) as a function of $\alpha$, $\gamma$, $\kappa$, and $\mu$ (in (kJ/mol)$^2$).
In each contour plot, $C_{i,j}$ is given as a function of $\kappa$ and $\mu$, whereas $\alpha$ and $\gamma$ are kept constant.
}
\label{fig:param4}
\end{figure}

To investigate whether the adjustment of $\kappa$ and $\mu$, in addition to $\alpha$ and $\gamma$, results in a significant increase in accuracy, the cost functions $C_{23,24}$ and $C_{23,25}$ depending on $\alpha$, $\gamma$, $\kappa$, and $\mu$ are given in Fig.~\ref{fig:param4} (results for additional structures are given in the Supporting Information).
In each contour plot the cost function depending on $\kappa$ and $\mu$ is given, whereby $\alpha$ and $\gamma$ are varied between contour plots.
Note that $\beta_c$ in the PBE functional depends on $\mu$, $\beta_c = 3 \mu / \pi^2$.
By comparing Figs.~\ref{fig:param4} (top) and (bottom), we see that for $\alpha=0.2$ and $\gamma=0.0$ the cost functions are similar.
From this result we conclude that the optimization of the parameters $\kappa$ and $\mu$, in addition to $\alpha$ and $\gamma$, is necessary to obtain a sufficiently flexible LC-PBE0 functional.

The four parameters were optimized employing the $D_\text{P}$ reference set and the following parameter values were obtained:
$\alpha = 0.176$ ($\sigma = 0.080$), $\gamma = 0.111$, $\kappa = 1.48$, and $\mu = 0.471$.
The functional with these parameters we refer to as LC*-PBE0($D_\text{P}$), where the star indicates that the original parameters were modified and `$D_\text{P}$' denotes that this parameter change was made for the $D_\text{P}$ reference data set.
All parameters clearly differ from the ones in LC-PBE0.
While the parameters $\kappa$ and $\mu$ were determined by fulfilling exact boundary conditions \cite{Perdew1996a}, the behavior of the functional between those boundary conditions may still be incorrect.
Hence, deviations from the exact parameters can lead to a functional that is more accurate for the chemical system of interest than LC-PBE0.
We emphasize that our LC$^\star$-PBE0 functional is system dependent in such a way that its optimum parameters will be different for different reference data sets.
However, this is not a drawback as the reliability of this class of functionals will be assessable according to
an error measure for each individual result in the error estimation procedure.

\subsection{Assessment of Re-Parametrization and Error Estimation}

Before we consider the conceptually decisive error estimation step for our system-dependent functionals, we first demonstrate that they in fact achieve a significant improvement with respect to accuracy for the reference data set.
While one might expect that this is naturally the case, it is not guaranteed because the explicit analytical form of the functional might not allow for 
such an improvement and the different reference data points might not be equally well representable by a common parameter set.

\begin{table}
\centering
\caption{Largest absolute deviation (LAD), mean absolute deviation (MAD), and mean signed deviation (MSD) of a selection of functionals, some with D3 dispersion corrections, for the $D_\text{P}$ reference set (in kJ/mol).}
\begin{tabular}{lrrr}
\hline
{} &    LAD &   MAD &    MSD \\
\hline
B3LYP     & 31.2 & 13.4 & -0.1 \\
B3LYP-D3  & 30.2 & 13.8 & -0.0 \\
BP86      & 65.0 & 33.1 & -8.6 \\
BP86-D3   & 66.5 & 35.5 & -8.6 \\
LC-PBE0   & 68.9 & 20.8 & -2.3 \\
M06-2X    & 69.6 & 28.1 &  4.6 \\
M06-2X-D3 & 69.6 & 28.1 &  4.7 \\
M06-L     & 45.7 & 24.7 & -1.6 \\
M06-L-D3  & 45.8 & 24.6 & -1.6 \\
PBE       & 66.3 & 32.8 & -8.1 \\
PBE0      & 32.3 & 13.6 &  0.1 \\
PBE0-D3   & 31.6 & 13.8 &  0.3 \\
TPSS      & 60.8 & 31.3 & -7.5 \\
TPSS-D3   & 62.2 & 32.9 & -7.4 \\
TPSSh     & 45.1 & 20.7 & -4.2 \\
TPSSh-D3  & 46.4 & 22.5 & -2.7 \\
\hline
LC$^\star$-PBE0($D_\text{P}$)  & 25.7 & 10.0 & -0.1 \\
\hline
\end{tabular}
\label{tab:mini_training}
\end{table}

In Table~\ref{tab:mini_training}, the accuracy of LC$^\star$-PBE0($D_\text{P}$) is compared to that of common density functionals (including D3 dispersion corrections).
LC$^\star$-PBE0($D_\text{P}$) features the lowest MAD, followed by B3LYP and PBE0.
As expected, GGA and meta-GGA functionals are less accurate than most hybrid functionals.
Moreover, due to the small molecular size, D3 corrections have no significant effect.
In addition, the MAD of no functional is within chemical accuracy and all functionals feature a high LAD of at least 25~kJ/mol.
Considering LC$^\star$-PBE0($D_\text{P}$) was fitted to this data set and still shows a LAD of 25.7~kJ/mol, underlines the fact that the electronic structure of transition metal complexes is difficult to reproduce by density functionals because of their restrictive functional form. 

While the results in Table~\ref{tab:mini_training} confirm the well-known fact \cite{Jiang2012} that density functionals applied to transition metal complexes rarely achieve chemical accuracy of about one kcal/mol, it is known that DFT can be very accurate for certain cases \cite{Goerigk2011a}. 
Clearly it is desirable to identify cases for which DFT fails and cases for which the results are reliable.

\begin{figure}
\centering
\includegraphics[width=\textwidth]{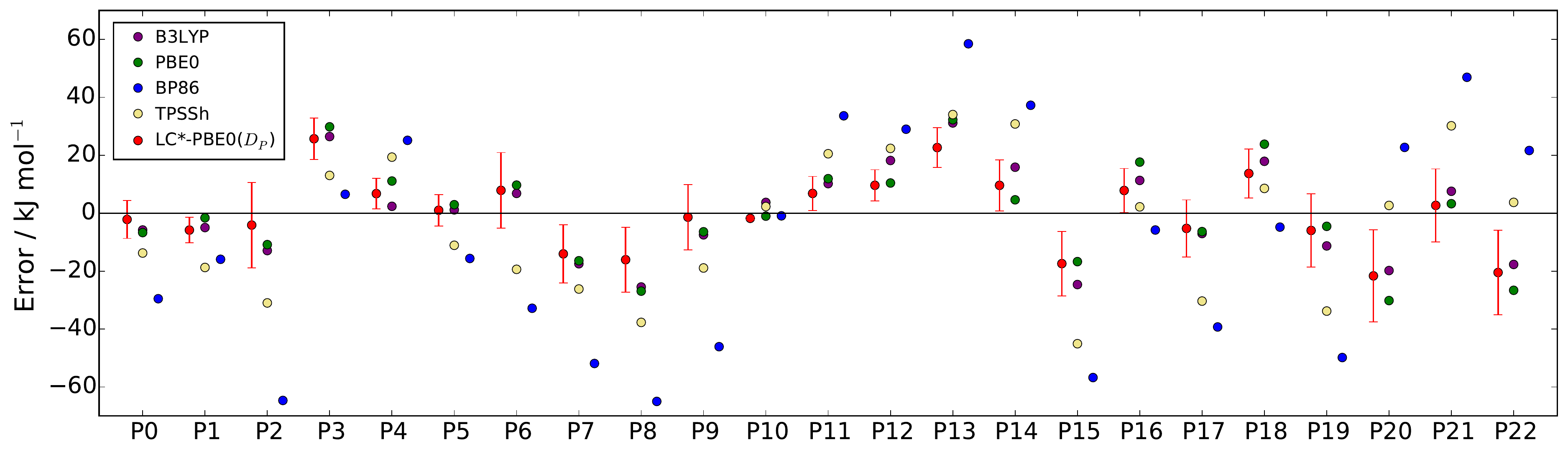}
\caption{
Errors of LC$^\star$-PBE0($D_\text{P}$) with error bars indicating a standard deviation and standard functionals for the reference data set $D_\text{P}$.
All data points in the set are denoted as P$i$.
}
\label{fig:mini_errors}
\end{figure}

As described in Section~\ref{subsec:bee} and \ref{subsec:deriv}, our functional allows for error estimates to be calculated.
With the standard deviation $\sigma$ and the best-fit parameters $a_0$, the normal distribution given in Eq.~(\ref{eq:normal_distribution}) can be sampled and a set of parameters $\vec{a} = \{a_1, a_2, \ldots, a_N\}$ can be generated (we introduce the vector notation to denote the set of parameter sets, which is a
set of parameters in this special case).
Employing the self-consistent electron density obtained from the functional with parameters $a_0$, the electronic energies for the parameters in $\vec{a}$ is calculated.
The standard deviation $\sigma(\mathcal{O}(i))$ is then calculated according to Eq.~(\ref{eq:standard_deviation}).
In Fig.~\ref{fig:mini_errors}, LC$^\star$-PBE0($D_\text{P}$) (with error bars, calculated from an ensemble of $N=25$ functionals given in the Supporting Information) is compared to popular density functionals with respect to $D_\text{P}$.
It can be seen that for most elements of the data set the error with respect to the reference is within one standard deviation.
For almost all reference data points the error is within two standard deviations; only for P3 and P13, the error was underestimated by LC$^\star$-PBE0($D_\text{P}$).

Further, the standard deviation reported by LC$^\star$-PBE0($D_\text{P}$) not always coincides with the spread of results from other functionals.
For example, the standard deviation of P4 is comparatively small (5.2~kJ/mol), whereas the errors of the other functionals is ranging from 2--34~kJ/mol.
Therefore, taking the spread of results from a set of functionals is not a stochastically meaningful indicator for the accuracy.

In addition, the errors of all functionals are highly unsystematic and the spread of errors is large.
This result is particularly striking, when considering the fact that the structures in our data set are homologous by construction.

\subsection{Transferability of the Model System}

For the reference data set $D_\text{P}$, we showed that the re-parameterization of the LC-PBE0 resulted in a significantly more accurate 
functional LC$^\star$-PBE0($D_\text{P}$), that also provides reliable error estimates for each result.
In this section, we investigate the transferability of the model system to the chemical system of interest.
As shown in Fig.~\ref{fig:pruning}, the (\textit{1-armed}) model which more closely resembles the core structure of the Yandulov--Schrock catalyst, probes the effect of the second coordination shell on the parameterization.

\begin{table}
\centering
\caption{Largest absolute deviation (LAD), mean absolute deviation (MAD), and mean signed deviation (MSD) of a selection of functionals, some with D3 dispersion corrections, for the $D_\text{A}$ reference set (in kJ/mol).}
\begin{tabular}{lrrr}
\hline
{} &  LAD &  MAD &  MSD \\
\hline
B3LYP     & 32.3 & 11.1 &  0.8 \\
B3LYP-D3  & 28.1 & 10.2 &  1.8 \\
BP86      & 70.1 & 24.8 & -8.1 \\
BP86-D3   & 68.0 & 25.8 & -7.0 \\
LC-PBE0   & 72.1 & 22.7 &  2.4 \\
M06-2X    & 71.1 & 25.9 &  6.1 \\
M06-2X-D3 & 71.1 & 25.8 &  6.0 \\
M06-L     & 50.2 & 17.5 & -5.0 \\
M06-L-D3  & 50.1 & 17.6 & -5.0 \\
PBE       & 71.5 & 24.5 & -8.7 \\
PBE0      & 31.9 & 12.7 & -0.6 \\
PBE0-D3   & 29.7 & 12.0 &  0.0 \\
TPSS      & 58.7 & 24.4 & -5.0 \\
TPSS-D3   & 56.8 & 25.1 & -4.2 \\
TPSSh     & 45.9 & 15.2 & -2.1 \\
TPSSh-D3  & 42.8 & 15.6 & -1.3 \\
\hline
LC$^\star$-PBE0($D_\text{P}$)  & 23.3 &  8.7 & 0.0 \\
LC$^\star$-PBE0($D_\text{A}$)  & 20.8 &  7.2 & 0.1 \\
\hline
\end{tabular}
\label{tab:1arm}
\end{table}

In Table~\ref{tab:1arm}, the accuracy of LC$^\star$-PBE0($D_\text{P}$) and popular density functionals (some including D3 dispersion corrections) with respect to the data set $D_\text{A}$ is shown.
With an MAD of $8.7$~kJ/mol, LC$^\star$-PBE0($D_\text{P}$) is more accurate than all other standard functionals.
Furthermore, due to increased system size, the contribution of the D3 corrections rose compared to $D_\text{P}$ and has a slight positive effect on the MAD for most functionals.
Finally, the strikingly high LAD of density functionals with a reasonable MAD (e.g., B3LYP-D3), highlights the need for a method with error estimation.

To investigate the effect of the model system on the parameterization, the parameters of LC$^\star$-PBE0 were optimized for $D_\text{A}$ to yield LC$^\star$-PBE0($D_\text{A}$).
The obtained optimal parameters are: 
$\alpha = 0.128$ ($\sigma = 0.081$), $\gamma = 0.080$, $\kappa = 1.49$, and $\mu = 0.512$.
In comparison to the parameters of LC$^\star$-PBE0($D_\text{P}$), only $\alpha$ and $\gamma$ changed, whereas $\kappa$ and $\mu$ remained more or less the same.
From Table~\ref{tab:1arm}, it can be seen that also the LAD and MAD decreased only slightly compared to LC$^\star$-PBE0($D_\text{P}$).
This suggests that it is the flexibility of the functional and not the choice of the model system that limits its accuracy.

\begin{figure}
\centering
\includegraphics[width=1\textwidth]{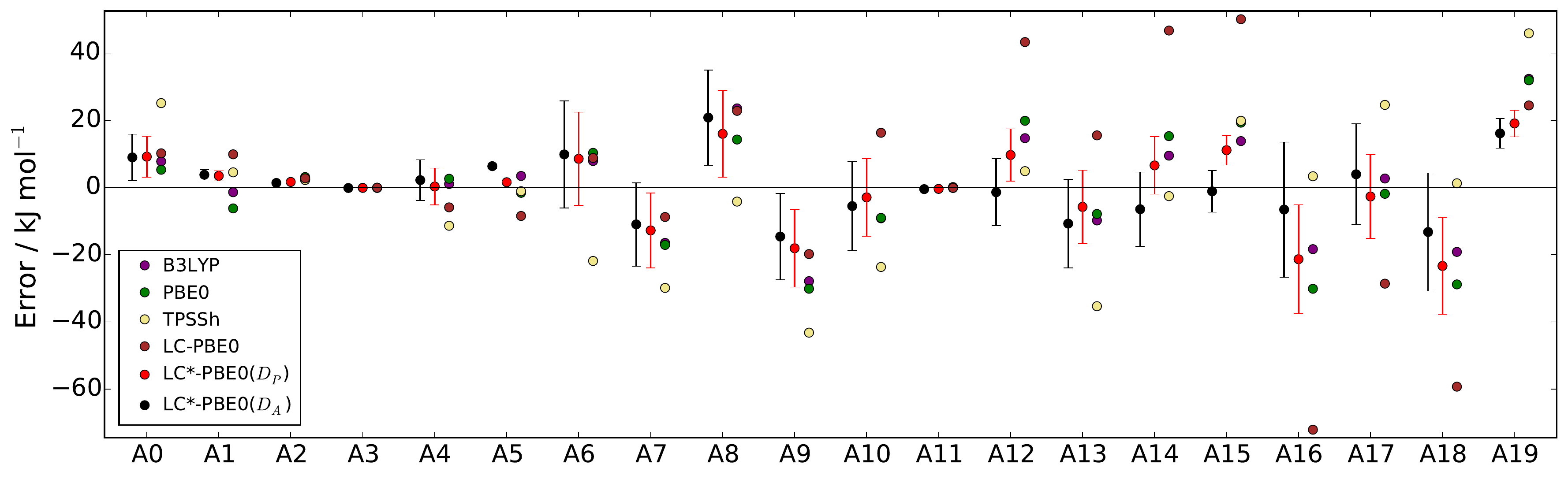}
\caption{Errors of LC$^\star$-PBE0($D_\text{P}$) and LC$^\star$-PBE0($D_\text{A}$) (with error bars indicating $\pm 1$ standard deviation) and standard functionals for data set $D_\text{A}$.
All data points in the set are denoted as A$i$.
}
\label{fig:1armed_errors}
\end{figure}

In Fig.~\ref{fig:1armed_errors}, the errors of LC$^\star$-PBE0($D_\text{P}$), LC$^\star$-PBE0($D_\text{A}$), and standard density functionals with respect to $D_\text{A}$ are shown.
It can be seen that the error bars reported by both error estimation functionals give a reliable and consistent indication for the accuracy of a result: in nearly all cases the actual error is within two standard deviations.

\section{Error Estimation for the Chatt--Schrock Cycle}

In the following section, the LC$^\star$-PBE0($D_\text{A}$) functional is applied to study reaction energies in the Chatt--Schrock cycle.
The energetics of this cycle were subjected to many theoretical studies \cite{LeGuennic2005,Reiher2005,Studt2005,Magistrato2007,Schenk2008,Schenk2009,Schenk2009a,Thimm2015,Bergeler2015b,Bergeler2015}.
Due to different computational setups (e.g., model catalyst, density functional, and basis sets), the results of these studies varied.
In Table~\ref{tab:chatt_schrock}, the calculated reaction energies for the complete Chatt--Schrock cycle including standard deviations are given.
While the majority of reactions features a small standard deviation of below 6 kJ/mol, there are reactions for which the functional predicts an unacceptably large error.
For example, with a standard deviation of 18.7~kJ/mol the reaction energy of the first protonation is apparently difficult to determine, whereas
LC$^\star$-PBE0($D_\text{A}$) reports a low uncertainty for subsequent protonation reactions.

\begin{table}
\centering
\caption{LC$^\star$-PBE0($D_\text{A}$) reaction energies (with standard deviations) for the first and second half of the full Chatt--Schrock cycle in kJ/mol.
LutH$^+$ and CrCp$^*_2$ are abbreviated as AH$^+$ and R, respectively.}
\begin{tabular}{lrr}
\hline
Reaction & $\Delta E$ &  $\sigma$ \\
\hline
$[$Mo$]$-N$_2$ + AH$^+$ $\rightarrow$ \{[Mo]-N$_2$H\}$^+$ + A     &      27.8 &   18.7 \\
\{[Mo]-N$_2$H\}$^+$ + R $\rightarrow$ [Mo]-N$_2$H + R$^+$         &    -120.9 &    5.9 \\
$[$Mo$]$-N$_2$H + AH$^+$ $\rightarrow$ \{[Mo]-N$_2$H$_2$\}$^+$ + AH   &    -103.4 &    2.6 \\
\{[Mo]-N$_2$H$_2$\}$^+$ + R $\rightarrow$ [Mo]-N$_2$H$_2$ + R$^+$     &      21.8 &   10.6 \\
$[$Mo$]$-N$_2$H$_2$ + AH$^+$ $\rightarrow$ \{[Mo]-N$_2$H$_3$\}$^+$ + AH &     -40.0 &    6.1 \\
\{[Mo]-N$_2$H$_3$\}$^+$ + R $\rightarrow$ [Mo]-N$_2$H$_3$ + R$^+$     &    -237.7 &    5.3 \\
\hline
$[$Mo$]$-N + AH$^+$ $\rightarrow$ \{[Mo]-NH\}$^+$ + A         &     -74.4 &    5.4 \\
\{[Mo]-NH\}$^+$ + R $\rightarrow$ [Mo]-NH + R$^+$             &       0.2 &   10.5 \\
$[$Mo$]$-NH + AH$^+$ $\rightarrow$ \{[Mo]-NH$_2$\}$^+$ + AH       &    -151.8 &    1.7 \\
\{[Mo]-NH$_2$\}$^+$ + R $\rightarrow$ [Mo]-NH$_2$ + R$^+$         &     -22.7 &   15.1 \\
$[$Mo$]$-NH$_2$ + AH$^+$ $\rightarrow$ \{[Mo]-NH$_3$\}$^+$ + AH     &    -146.7 &    1.1 \\
\{[Mo]-NH$_3$\}$^+$ + R $\rightarrow$ [Mo]-NH$_3$ + R$^+$         &       9.4 &    3.0 \\
$[$Mo$]$-NH$_3$ + N$_2$ $\rightarrow$ [Mo]-N$_2$ + NH$_3$         &      -7.6 &   13.6 \\
\hline
\end{tabular}
\label{tab:chatt_schrock}
\end{table}

Since the parameters in LC$^\star$-PBE0($D_\text{A}$) were optimized for a data set which contains neither the reducing agent CrCp$^*_2$ nor the acid lutidinium, no error can be calculated for either the oxidation of CrCp$^*_2$ or for the abstraction of the proton from lutidinium.
A more extensive data set needs to be constructed to be able to assign an uncertainty to these reactions.
We may therefore anticipate that the errors reported here underestimate the actual errors.
Since, however, the error of electron and proton abstraction would result in a constant shift for the reduction and protonation reactions, respectively, 
it does not affect our conclusions.

Due to the large HIPT substituents, calculations on the full Chatt--Schrock catalyst require dispersion corrections to be considered.
These cannot be well described by LC$^\star$-PBE0($D_\text{A}$) because $D_\text{A}$ does not contain reference data on large model complexes
for which dispersion is increasingly important. 
However, since no heptane solvent molecules are included in our Yandulov--Schrock structural models, dispersion corrections are not considered here as they
would artificially overestimate all intra-complex dispersion. Clearly, in general, dispersion corrections must be considered. As empirical force-field-type
dispersion corrections would require an extensive parametrization, we recommend density-based techniques (see, e.g., Refs.\ \cite{thakchenko,corminbeuf})
for a system-focused density functional optimization.

\begin{figure}
\centering
\includegraphics[width=\textwidth]{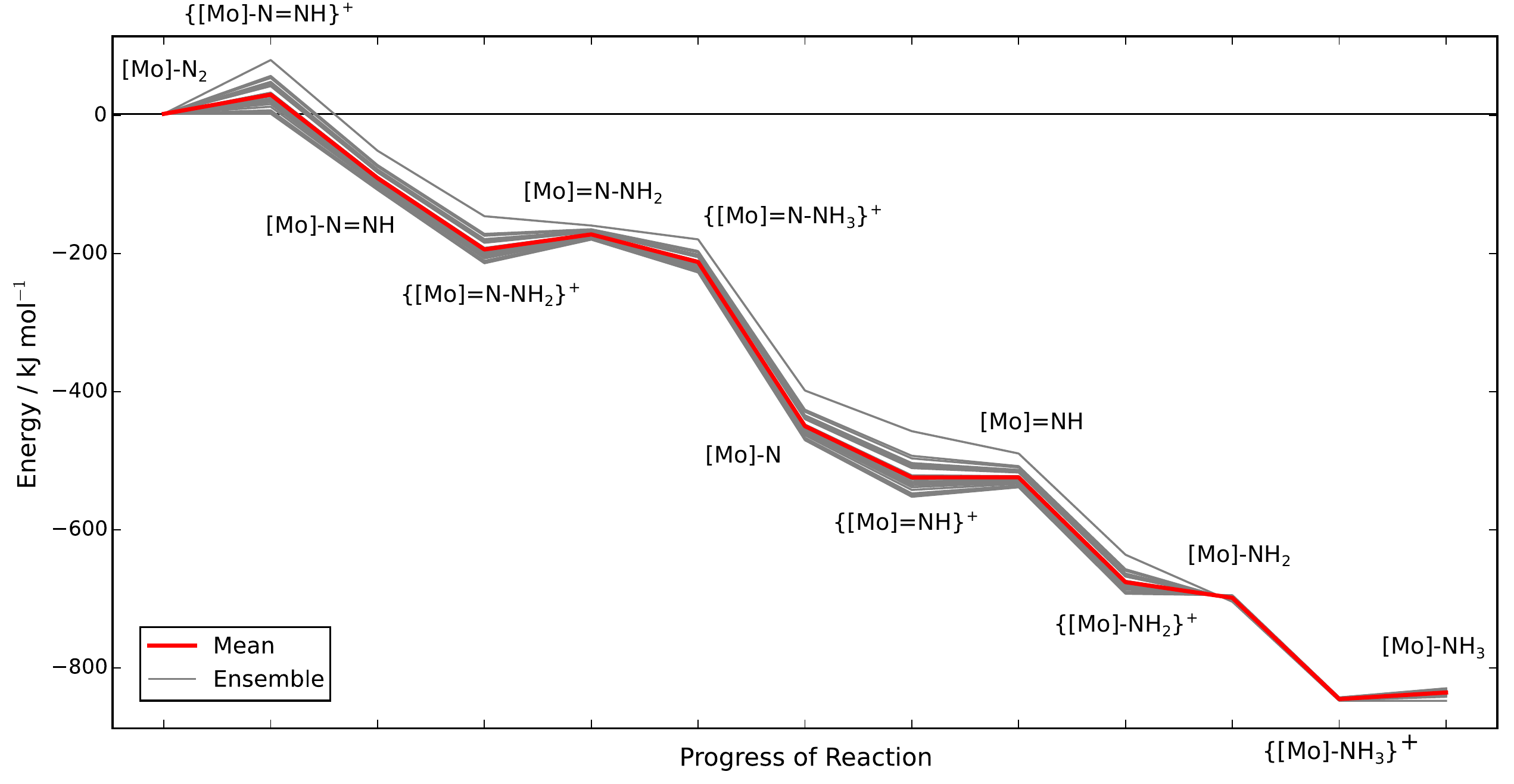}
\caption{Statistical representation of energy profile of Chatt-Schrock cycle.
Red: mean of LC$^\star$-PBE0($D_\text{A}$); gray: ensemble of LC$^\star$-PBE0($D_\text{A}$);}
\label{fig:sample_cycle}
\end{figure}

In Fig.~\ref{fig:sample_cycle}, the mean energy profile (red) together with the ensemble of LC$^\star$-PBE0($D_\text{A}$) (gray) is depicted.
The uncertainty associated with the energy of each intermediate with respect to the first intermediate of the cycle can be seen from the 
spread of the energy profiles.
Similarly, a change in spread of the energy profiles resembles the error of each reaction energy.
Fig.~\ref{fig:sample_cycle} highlights the importance of error estimation when interpreting reaction profiles commonly found in the literature.

\section{Conclusions}

In this work, a novel approach for the construction of reliable, system-specific density functionals with Bayesian error estimation is presented.
By employing a system-focused re-parametrization of the RSH functional LC-PBE0, we were able to obtain a functional that allows for the accurate description of a particular system of interest.
By choosing a functional based on physical principles with few parameters we also overcame the issue of transferability.
Whereas a system-specific parametrization of density functionals is in general not a recommended strategy, here it is viable and useful because our functional provides confidence intervals for each result, thereby allowing one to assess whether the reported result is reliable.
Clearly, our approach requires the generation of sufficiently accurate reference data for the class of molecules under consideration, but this is 
becoming comparatively easy with modern quantum chemistry software (see, e.g., Refs.\ \cite{Riplinger2013,Riplinger2016}) --- 
even for multi-configuration cases (see, e.g., Refs.\ \cite{keller,hedegard,Stein2016}).

We applied our approach to the Yandulov--Schrock catalyst and identified that parameters in both the long-range corrected scheme and the PBE functional need to be optimized to obtain a sufficiently flexible functional.
Furthermore, we were able to show that the reported error estimates are indeed reliable.
Finally, we calculated the reaction energies of the Chatt--Schrock cycle.
We showed that the confidence level of reaction energies can vary significantly
--- even if the reactions are very similar --- thus, highlighting the need for error estimation.

To further increase the functionals accuracy and error estimation reliability, a functional form with greater flexibility would be beneficial, 
which is currently investigated in our laboratory.

\section*{Acknowledgments}
This work has been financially supported by the Schweizerischer Nationalfonds. 
GNS gratefully acknowledges support by a fellowship of the Fonds der Chemischen Industrie.



\providecommand{\url}[1]{\texttt{#1}}
\providecommand{\urlprefix}{}
\providecommand{\foreignlanguage}[2]{#2}
\providecommand{\Capitalize}[1]{\uppercase{#1}}
\providecommand{\capitalize}[1]{\expandafter\Capitalize#1}
\providecommand{\bibliographycite}[1]{\cite{#1}}
\providecommand{\bbland}{and}
\providecommand{\bblchap}{chap.}
\providecommand{\bblchapter}{chapter}
\providecommand{\bbletal}{et~al.}
\providecommand{\bbleditors}{editors}
\providecommand{\bbleds}{eds.}
\providecommand{\bbleditor}{editor}
\providecommand{\bbled}{ed.}
\providecommand{\bbledition}{edition}
\providecommand{\bbledn}{ed.}
\providecommand{\bbleidp}{page}
\providecommand{\bbleidpp}{pages}
\providecommand{\bblerratum}{erratum}
\providecommand{\bblin}{in}
\providecommand{\bblmthesis}{Master's thesis}
\providecommand{\bblno}{no.}
\providecommand{\bblnumber}{number}
\providecommand{\bblof}{of}
\providecommand{\bblpage}{page}
\providecommand{\bblpages}{pages}
\providecommand{\bblp}{p}
\providecommand{\bblphdthesis}{Ph.D. thesis}
\providecommand{\bblpp}{pp}
\providecommand{\bbltechrep}{Tech. Rep.}
\providecommand{\bbltechreport}{Technical Report}
\providecommand{\bblvolume}{volume}
\providecommand{\bblvol}{Vol.}
\providecommand{\bbljan}{January}
\providecommand{\bblfeb}{February}
\providecommand{\bblmar}{March}
\providecommand{\bblapr}{April}
\providecommand{\bblmay}{May}
\providecommand{\bbljun}{June}
\providecommand{\bbljul}{July}
\providecommand{\bblaug}{August}
\providecommand{\bblsep}{September}
\providecommand{\bbloct}{October}
\providecommand{\bblnov}{November}
\providecommand{\bbldec}{December}
\providecommand{\bblfirst}{First}
\providecommand{\bblfirsto}{1st}
\providecommand{\bblsecond}{Second}
\providecommand{\bblsecondo}{2nd}
\providecommand{\bblthird}{Third}
\providecommand{\bblthirdo}{3rd}
\providecommand{\bblfourth}{Fourth}
\providecommand{\bblfourtho}{4th}
\providecommand{\bblfifth}{Fifth}
\providecommand{\bblfiftho}{5th}
\providecommand{\bblst}{st}
\providecommand{\bblnd}{nd}
\providecommand{\bblrd}{rd}
\providecommand{\bblth}{th}

%
%
%
%
%
%

\end{document}